# Density function associated with nonlinear bifurcating map

A. D. Alhaidari

*Shura Council, Riyadh 11212, Saudi Arabia*

E-mail: haidari@mailaps.org

In the class of nonlinear one-parameter real maps we study those with bifurcation that exhibits period doubling cascade. The fixed points of such a map form a finite discrete real set with dimension $2^n m$, where *m* is the (odd) number of "primary branches" of the map in the non-chaotic region and *n* is a non-negative integer. We associate with this map a nonlinear dynamical system whose Hamiltonian matrix is real, tridiagonal and symmetric. The density of states of the system is calculated and shown to have a number of separated bands equals to $2^{n-1} m$ for $n \neq 0$, in which case the density has *m* bands. The location of the bands depends only on the map parameter and the odd/even ordering of the fixed points in the set. Polynomials orthogonal with respect to this density (weight) function are constructed. The logistic map is taken as an illustrative example.

## I. INTRODUCTION

For a given real parameter $\lambda$ we consider the nonlinear map $f_\lambda$ whose infinite action takes a segment of the real line, $x \in [x_-, x_+]$, into a finite dimensional real discrete set. This is the set of stable fixed points of the map, which is parameterized by $\lambda$. The dimension of the set depends on the range of values to which $\lambda$ belongs. However, for other values of $\lambda$, such maps exhibit chaotic behavior where the limit of the map does not exist but, usually, bounded. In the present work we study these maps in the non-chaotic region. Typically, this class of nonlinear maps bifurcate at a given set of values of the parameter $\lambda$, called the bifurcation points (a.k.a. branch points). We restrict our investigation to those maps with bifurcation diagrams that consist of period doubling cascade. In other words, the bifurcation diagram (i.e., the set of fixed points of the map for all values of $\lambda$) consists of branches that bifurcate simultaneously (at the same value of $\lambda$) into two branches and for several values of $\lambda$.

This subclass of nonlinear one-parameter real maps has been studied extensively in the mathematics and physics literature. For an introductory review one may consult any one of the large number of books on the subject. The list shown in Ref. [1] is only an example of those that are of relevance to our present work. Our contribution here is to give a new and alternative realization of such maps that should give more insight into the physical interpretation and enhance their applications in mathematical physics and nonlinear dynamics. To each map $f_\lambda$ in this subclass and for a given initial point (state) $x_0$ that belongs to its domain $[x_-, x_+]$ we associate a real symmetric matrix which is tridiagonal by construction and parameterized by $\lambda$ and $x_0$. Consequently, one could interpret such hermitian matrix as the matrix representation (in some complete basis) of the Hamiltonian of a dynamical system that corresponds to the map. The density of states of the system could thus be calculated using any one of the techniques that were developed for tridiagonal Hamiltonian matrices [2]. We will show that this density has a band structure with forbidden gaps. The number of the density bands equals half of the dimension of the set of fixed points of the map. Whereas, the location of the bands are independent of $x_0$; they only depend on $\lambda$ and the ordering (odd or even) of the set of fixed points. Additionally, we use the theory of orthogonal polynomials and their three-term recursion relations associated with tridiagonal matrices to find the set of polynomials orthogonal with respect to this density (weight) function.

In the following section we start by presenting the mathematical notation necessary for the definition of objects needed to establish our findings. The tridiagonal symmetric matrix associated with $f_\lambda$ and $x_0$ is constructed. The corresponding density function will be evaluated using the resolvent operator of the system. The "logistic map" is taken as an illustrative example and the density of states with single and multiple bands will be given. Polynomials that are orthogonal with respect to these density (weight) functions are constructed in Sec. III.

## II. TRIDIAGONAL HAMILTONIAN AND THE DENSITY OF STATES

Let $x_0 \in [x_-, x_+]$ and define $x_{n+1} = f_\lambda(x_n)$ for $n = 0,1,2,\ldots$ Therefore, the set of fixed points of the map in a non-chaotic region is given by the finite multi-valued limit,

$$\lim_{n\to\infty} x_n = \{c_k(\lambda)\}_{k=1}^N . \tag{0}$$

The dimension of this set, *N*, and the values of its elements are independent of the initial seed $x_0$. They depend only on the real parameter $\lambda$ [1]. Let *m* be the number of "primary branches" in the bifurcation diagram for the given non-chaotic region (i.e., the branches at the minimum value of $\lambda$ in that region). Therefore, if the bifurcation period doubling process in the region takes place at the bifurcation points $\lambda = \lambda_1, \lambda_2, \lambda_3, ..$ etc., then the number of branches in the cascade will be $m, 2m, 4m, ..$ etc. Thus, the dimension of the set of fixed points of the map in the given non-chaotic region is $N = 2^n m$ for all $\lambda$ in the range $\lambda_n < \lambda < \lambda_{n+1}$. Schematically, this process could be written as

$$x_0 \xrightarrow{f_\lambda} x_1 \xrightarrow{f_\lambda} x_2 \xrightarrow{f_\lambda} \ldots \xrightarrow{f_\lambda} \left\{ \begin{matrix} c_1 \\ \downarrow \\ c_2 \\ \downarrow \\ \vdots \\ c_N \end{matrix} \right\} \tag{1}$$

where the limiting action of the map cycles through the set of *N* fixed points $\{c_k\}_{k=1}^N$. As an illustration, we consider the well-known logistic map where $f_\lambda(x) = \lambda x(1-x)$, $x \in [0,1]$, and $0 \le \lambda \le 4$. Figure 1 shows the bifurcation diagram of the map in the non-chaotic region $2.8 < \lambda < 3.6$ with one primary branch (i.e., *m* = 1). The period doubling cascade is evident as well as the first few bifurcation points, $\lambda_i$ = 3.00, 3.45, 3.54, 3.56. The process, which is given schematically in (1), is shown graphically in Fig. 2 for the logistic map in the region with period-4 orbits and where $\lambda$ = 3.5. It is also known that the logistic map has another non-chaotic region with period three (*m* = 3) doubling cascade in the neighbor-hood of $\lambda \ge$ 3.828.

Now if we define the set of real numbers $a_n = x_{2n}$ and $b_n = x_{2n+1}$ for $n = 0,1,2,...$ then we could construct the following real symmetric tridiagonal matrix

$$H_{x_0}^{\lambda} = \begin{pmatrix} a_0 & b_0 & & & & & \\ b_0 & a_1 & b_1 & & \mathbf{0} & & \\ & b_1 & a_2 & b_2 & & & \\ & & b_2 & \times & \times & & \\ & & & \times & \times & \times & \\ & \mathbf{0} & & & \times & \times & \times \\ & & & & & \times & \times \end{pmatrix} \qquad (2)$$

This construction could be displayed in away that clearly show the successive action of the map $f_\lambda$ on $x_0$ as

$H_{x_0}^{\lambda} =$ 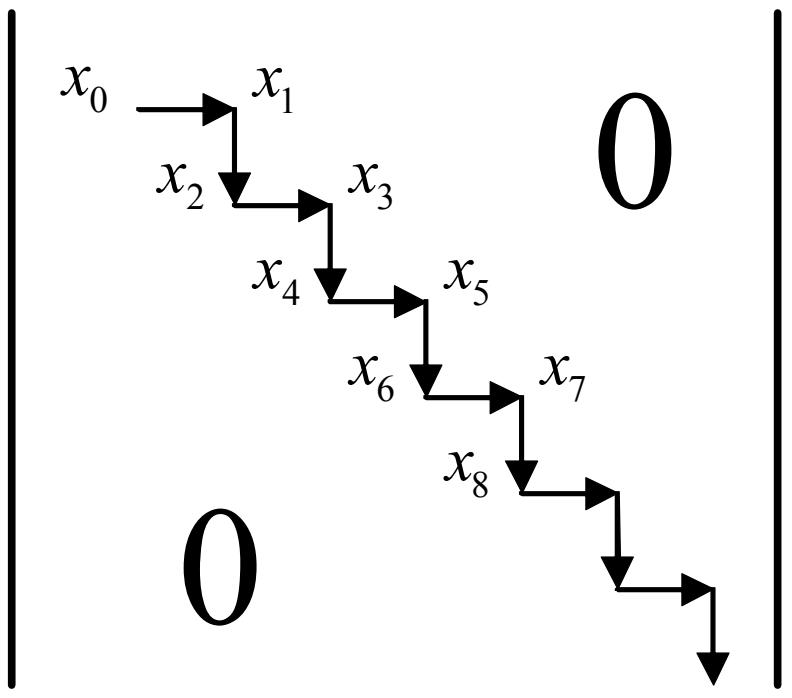 (2)’

The infinitely far tail of this matrix will be made up of a repeated sequence (of length $N/2$) containing the set of fixed point of the map. That is,

$$\lim_{n\to\infty} a_n = \left\{c_{2i-1}(\lambda)\right\}_{i=1}^{N/2} \equiv \left\{\alpha_i\right\}_{i=1}^{N/2}, \ \lim_{n\to\infty} b_n = \left\{c_{2i}(\lambda)\right\}_{i=1}^{N/2} \equiv \left\{\beta_i\right\}_{i=1}^{N/2}. \qquad (3)$$

However, if the order of the fixed points in the set, $\left\{c_k(\lambda)\right\}_{k=1}^{N}$, is changed by a one-element shift to the right or left then $\alpha_i$ and $\beta_i$ will be exchanged. We refer to the former ordering as even and to the latter as odd. The hermitian matrix (2) is a faithful representation of the nonlinear map $f_\lambda$ at $x_0$ since it contains all information about the action of the map at that point. On the other hand, $H_{x_0}^{\lambda}$ could also be interpreted as the matrix representation of the Hamiltonian of a corresponding nonlinear system in the non-chaotic region. Changing $x_0$ will not affect the far tail of the tridiagonal matrix $H_{x_0}^{\lambda}$ except, possibly, for the ordering ambiguity. The resolvent operator (Green’s function) associated with $H_{x_0}^{\lambda}$ is formally written as $G_{x_0}^{\lambda}(z) = \left(H_{x_0}^{\lambda} - z\right)^{-1}$, which is an analytic function in the complex *z*-plane except at the set of eigenvalues (discrete and continuous) of $H_{x_0}^{\lambda}$. One realization of the (0,0) component of this resolvent operator is given by the following infinite continued fraction [3,4]

$$G_{x_0}^{\lambda}(z) = \cfrac{-1}{z - a_0 - \cfrac{b_0^2}{z - a_1 - \cfrac{b_1^2}{z - a_2 - ...}}} \qquad (4)$$

where, for simplicity of notation, we have omitted the (0,0) subscript on $G_{x_0}^{\lambda}$. This continued fraction could be approximated, for a given large enough integer *M*, as

$$G_{x_0}^{\lambda}(z) \cong \cfrac{-1}{z-a_0-\cfrac{b_0^2}{z-a_1-\cfrac{b_1^2}{z-a_2-\ldots-\cfrac{b_{M-2}^2}{z-a_{M-1}-T(z)}}}} \quad (5)$$

where the function $T(z)$, called the "terminator," is written as

$$T(z) = \cfrac{\beta_1^2}{z-\alpha_1-\cfrac{\beta_2^2}{z-\alpha_2-\ldots-\cfrac{\beta_{N/2}^2}{z-\alpha_{N/2}-T(z)}}} \quad (6)$$

$\alpha_i$ and $\beta_i$ are given by Eq. (3) in terms of the fixed points of the map at $\lambda$. A measure of the accuracy of the approximation (5) is given by how small are the deviations $|a_M|-|\alpha_1|$ and $|b_{M-1}|-|\beta_1|$. The integer $M$ could, thus, be increase progressively to achieve the desired accuracy as well as calculation stability. The terminator $T(z)$ could easily by calculated using Eq. (6) once the set of coefficients $\alpha_i$ and $\beta_i$ are determined by the fixed points. In fact, for small values of $N$, an analytic expression for $T(z)$ could be obtained with little effort. As an example, for $m = 1$ and $N = 2$, we obtain

$$T(z) = \tfrac{1}{2}(z-\alpha_1) - \tfrac{1}{2}\sqrt{(z-\alpha_1-2\beta_1)(z-\alpha_1+2\beta_1)}\,. \quad (7)$$

The density of states $\rho_{x_0}^{\lambda}(y)$ of the dynamical system whose Hamiltonian is given by Eq. (2) could be obtained using the Green's function $G_{x_0}^{\lambda}(z)$ as [2,5,6]

$$\rho_{x_0}^{\lambda}(y) = \lim_{\varepsilon\to 0} \tfrac{1}{2\pi \mathrm{i}}\left[G_{x_0}^{\lambda}(y+\mathrm{i}\varepsilon) - G_{x_0}^{\lambda}(y+\mathrm{i}\varepsilon)\right] = \tfrac{1}{\pi}\,\mathrm{Im}\left[G_{x_0}^{\lambda}(y)\right], \quad (8)$$

where $y \in [y_-, y_+]$. Thus, the density is a measure of the discontinuity of the resolvent operator across the cut along the real line (named $y$ because we have already used $x$ to refer to the domain of the map) in the complex $z$-plane. This expression shows that the density vanishes wherever the value of the resolvent (on the real line) is real. Merging this fact with the approximation (5) we also conclude that the density vanishes wherever the value of the terminator $T(y)$ is real. Now, it is obvious that (6) gives a quadratic equation in $T(y)$. The solution of this equation for $T(y)$ results in a discriminant (the expression under the square root) which is a polynomial of degree $N$ in $y$ with real and $\lambda$-dependent coefficients. The hermiticity of the Hamiltonian matrix and its asymptotic structure dictates that this polynomial has $N$ real distinct roots, $y_1 < y_2 < \ldots < y_N$ [3,4,7]. They are independent of $x_0$ and depend only on $\lambda$. Moreover, the value of this polynomial (the discriminant) between these roots alternate between positive and negative values. Thus, the value of $T(y)$ between these roots alternate between real and complex values. Consequently, the density structure will consist of $\frac{N}{2}$ non-zero bands interleaved with $\frac{N}{2}-1$ forbidden gaps where the density vanishes. That is,

$$\rho_{x_0}^{\lambda}(y) = 0\text{, for all } y \text{ in } y_{2i} < y < y_{2i+1}, \quad (9)$$

where $i = 1, 2, \ldots, \frac{N}{2}-1$, $y_- = y_1$ and $y_+ = y_N$. As a simple example, where $N = 2$, Eq. (7) gives the two roots $y_{\pm} = \alpha_1 \pm 2\beta_1$. The density of states has a single band with non-zero support between these two roots while vanishing outside. For $N = 4$, Eq. (6) results in four real roots as follows:

$$\frac{1}{2}(\alpha_1+\alpha_2)+\frac{1}{2}\sqrt{(\alpha_1-\alpha_2)^2+4(\beta_1\pm\beta_2)^2}\ , \qquad (10.1)$$
$$\frac{1}{2}(\alpha_1+\alpha_2)-\frac{1}{2}\sqrt{(\alpha_1-\alpha_2)^2+4(\beta_1\pm\beta_2)^2}\ . \qquad (10.2)$$
Thus, the density function is made up of two bands with a gap in between. The analytic expressions of the roots become much more complicated for higher values of $N$ where one has to resort to numerical means to calculate them.

In the following graphical examples we use the logistic map and work in the non-chaotic region where the bifurcation diagram (shown in Fig. 1) exhibits period doubling cascade. Additionally, the accuracy and stability of the density calculation are with respect to the choice $M$ = 100 in Eq. (5). Figure 3a shows a single band density of the system before it gets into the cascade region and where $\lambda$ = 2.5. On the other hand, Figure 3b gives another single band density for a given initial state $x_0$ but the system is now in the period-2 bifurcation orbit where $\lambda_1<\lambda=3.2<\lambda_2$. Figure 4a shows two double band densities for two different $x_0$'s but with the same $\lambda$ = 3.5. The system is in period-4 bifurcation orbit where $\lambda_2<\lambda<\lambda_3$ and the ordering of the set of fixed points is even. The same is repeated in Fig. 4b but for another two values of $x_0$ where the ordering of the same fixed points is odd. One should be able to observe the effect of the ordering on the location of the density bands due to the resulting parameter exchange $\alpha_i\leftrightarrow\beta_i$ as explained above. Figure 5 displays a four-band density associated with the system in the period-8 bifurcation orbit where $\lambda$ = 3.55. For the sake of completeness, we plot in Fig. 6 the density of states of the system in the non-chaotic region with period three orbits ($m$ = 3) for $\lambda$ = 3.835 and $x_0$ = 0.5.

In the following section we obtain the three-term recursion relation that defines the set of orthogonal polynomials associated with the non-linear map $f_\lambda$ at the given initial state $x_0$. The weight function of these polynomials is the density of states found above.

## III. ORTHOGONAL POLYNOMIALS ASSOCIATED WITH THE MAP

The eigenvalue equation for the Hamiltonian $H_{x_0}^{\lambda}$ could be written as $H_{x_0}^{\lambda}|\psi\rangle$ = $y|\psi\rangle$. We expand the eigenvector $|\psi\rangle$ in a complete basis with the Fourier expansion coefficients $\{d_n(y)\}_{n=0}^{\infty}$. Assuming that the basis is chosen such that the matrix representation of the Hamiltonian $H_{x_0}^{\lambda}$ is tridiagonal and given by Eq. (2), then the eigenvalue equation above results in the following three-term recursion relation for the expansion coefficients
$$y\,d_n(y)=a_n d_n(y)+b_{n-1}d_{n-1}(y)+b_n d_{n+1}(y),\ n\geq 1 \qquad (11)$$
with the initial relation ($n$ = 0): $(a_0-y)\,d_0(y)+b_0 d_1(y)=0$. The solution of this recursion relation is written in terms of orthogonal polynomials [7,8]. We refer to these polynomial solutions as $\left\{P_n^{(\lambda,x_0)}(y)\right\}_{n=0}^{\infty}$ and normalize them by taking $P_0^{(\lambda,x_0)}(y)=1$. Therefore, all of them become well-defined just from knowledge of the recursion coefficients $\{a_n,b_n\}_{n=0}^{\infty}$, which are the matrix elements of the Hamiltonian. That is, we can write them as

$$
\begin{aligned}
&P_0^{(\lambda,x_0)}(y)=1\\
&P_1^{(\lambda,x_0)}(y)=\tfrac{y-a_0}{b_0}\\
&P_2^{(\lambda,x_0)}(y)=\tfrac{1}{b_1}\left[\left(y-a_1\right)P_1^{(\lambda,x_0)}(y)-b_0\right]\\
&P_3^{(\lambda,x_0)}(y)=\tfrac{1}{b_2}\left[\left(y-a_2\right)P_2^{(\lambda,x_0)}(y)-b_1P_1^{(\lambda,x_0)}(y)\right]\\
&P_4^{(\lambda,x_0)}(y)=\tfrac{1}{b_3}\left[\left(y-a_3\right)P_3^{(\lambda,x_0)}(y)-b_2P_2^{(\lambda,x_0)}(y)\right]\\
&\ldots.
\end{aligned}
\tag{12}
$$

The theory of orthogonal polynomials associated with tridiagonal symmetric matrices show that the weight function for these polynomials is the density function $\rho_{x_0}^{\lambda}(y)$ obtained above [3,4]. That is, we can write the orthogonality relations as

$$\int_{y_-}^{y_+} \rho_{x_0}^{\lambda}(y)P_n^{(\lambda,x_0)}(y)P_m^{(\lambda,x_0)}(y)dy=\delta_{nm}\,. \tag{13}$$

The integral breaks up into $N/2$ segments corresponding to the non-zero support (bands) of the density function. Figure 7 is a plot of the first few orthogonal polynomials $\left\{P_n^{(3.5,0.3)}(y)\right\}_{n=3}^{7}$ associated with the non-linear logistic map in the period-4 bifurcation orbit.

Polynomials of the second kind, which we refer to by $\left\{Q_n^{(\lambda,x_0)}(y)\right\}_{n=0}^{\infty}$, satisfy the same recursion relation (11). However, their corresponding initial relation ($n$ = 0) is not the same. It is non-homogeneous and reads: $\left(a_0-y\right)Q_0(y)+b_0Q_1(y)=1$. Moreover, $Q_0=0$. Thus, $Q_n^{(\lambda,x_0)}(y)$ is a polynomial with real coefficients and degree $n-1$, whereas $P_n^{(\lambda,x_0)}(y)$ is of degree $n$. Additionally, one can use these two kinds of polynomials to write an alternative representation of the (0,0) component of the resolvent operator as follows [3,4]

$$G_{x_0}^{\lambda}(z)=-\lim_{n\to\infty}\left[Q_n^{(\lambda,x_0)}(z)/P_n^{(\lambda,x_0)}(z)\right]. \tag{14}$$

In conclusion, we summarize our findings as follows. In the class of nonlinear one-parameter real maps we considered those with bifurcation diagrams that exhibit period doubling cascade. We associated with such map a real tridiagonal symmetric matrix parameterized by the initial state $x_0$ and the map parameter $\lambda$. Interpreting this as the Hamiltonian matrix of a corresponding nonlinear dynamical system establishes a one-to-one correspondence between the nonlinear map and the density of states of the system. Our study showed that the density has a band structure. The number and location of these bands depended on the number and values of the fixed points of the map. They are also invariant with respect to variations in $x_0$, except for a possible ambiguity in the ordering of the fixed points in the set. The logistic map was taken as illustrative example where several density of state functions where given. Moreover, the orthogonal polynomials of the first and second kind associated with the nonlinear map were also constructed. It is hoped that this study leads to more insight into the physical interpretation of such maps and enhance their applications in mathematical physics and nonlinear dynamics.

## ACKNOWLEDGMENTS

I'm grateful to H. A. Yamani and M. S. Abdelmonem for motivating discussions and to H. Bahlouli for the support in literature survey.

**FIGURES CAPTION:**

Fig. 1: The bifurcation diagram of the logistic map in the non-chaotic region $2.8 < \lambda < 3.6$. The period doubling cascade is evident as well as the first few branch points at $\lambda = 3.00, 3.45, 3.54, 3.56$

Fig. 2: The repeated action of the logistic map on the initial point $x_0 = 0.3$ in the region with period-4 orbits and where $\lambda = 3.5$. The four fixed point of the map are indicated by the horizontal dotted lines.

Fig. 3a: Single-band density associated with the map in the region before bifurcation. We took $\lambda = 2.5$ and $x_0 = 0.3$

Fig. 3b: Single-band density associated with the map in the period-2 bifurcation orbit. We took $\lambda = 3.2$ and $x_0 = 0.4$

Fig. 4a: Double-band densities for two different $x_0$'s but with the same $\lambda = 3.5$. The system is in period-4 bifurcation orbit and the ordering of the set of fixed points is even. The solid (dashed) curve corresponds to the initial point $x_0 = 0.30$ ($x_0 = 0.69$).

Fig. 4b: The same as Fig. 4a except for the ordering of the set of fixed points which is odd. The solid (dashed) curve corresponds to the initial point $x_0 = 0.15$ ($x_0 = 0.73$).

Fig. 5: Four-band density associated with the logistic map in the period-8 bifurcation orbit. We took $\lambda = 3.55$ and $x_0 = 0.3$. The edges of the four bands are located at $y = -1.267$, $-0.794$, $-0.743$, $0.339$, $0.546$, $1.640$, $1.665$, $2.150$

Fig. 6: The density of states of the system in the non-chaotic region with period-three orbit and for $\lambda = 3.835$, $x_0 = 0.5$. The edges of the three bands are located at $y = -0.783$, $-0.701$, $0.701$, $0.974$, $1.414$, $1.605$

Fig. 7 (color online): A plot of the orthogonal polynomials $\left\{P_n^{(\lambda,x_0)}(y)\right\}_{n=3}^{7}$ for $\lambda = 3.5$ and $x_0 = 0.5$ corresponding to the map in the period-two bifurcation orbit. The stems with cross head indicate the boundary of the non-zero support of the weight (density) function.

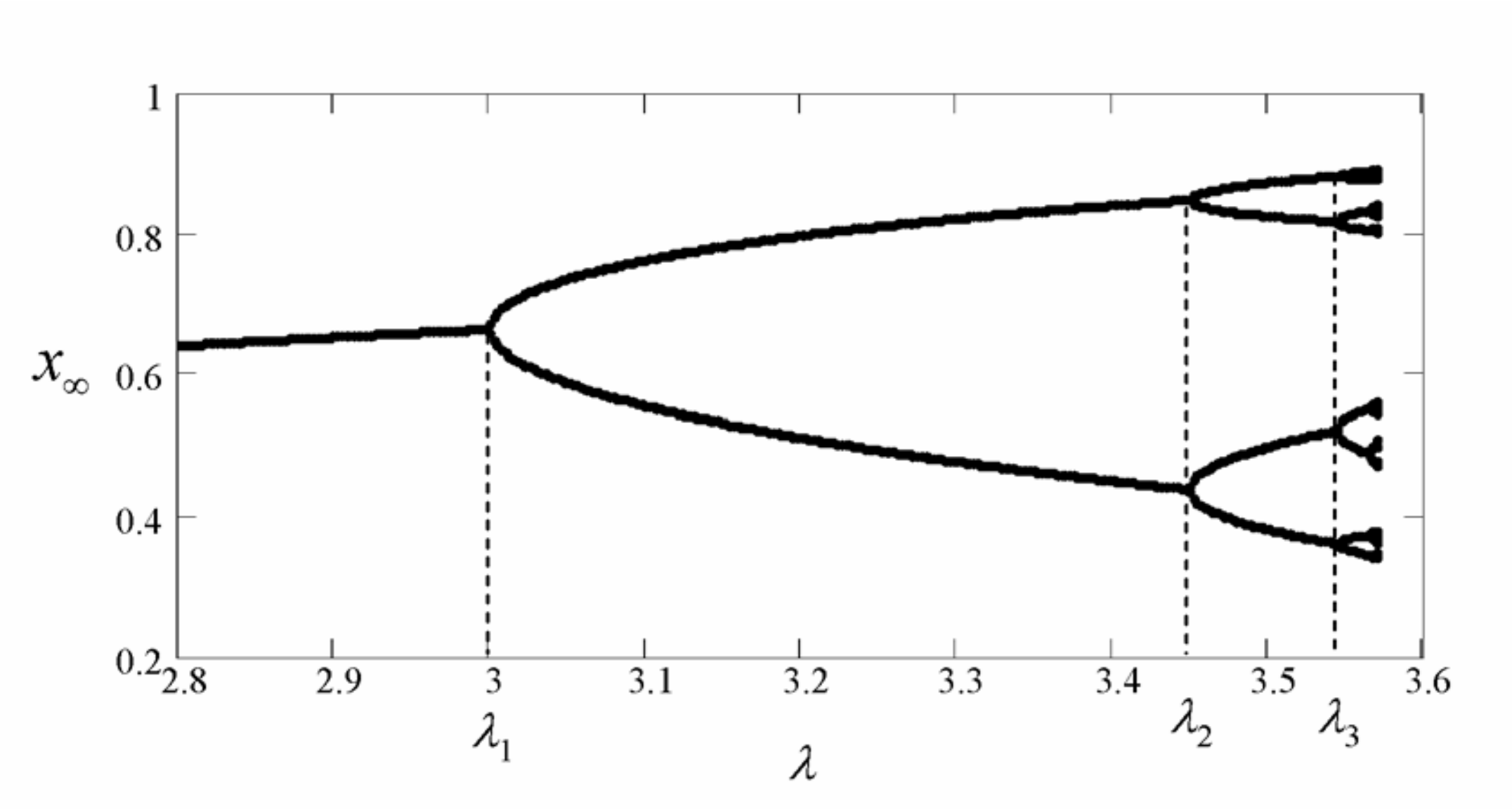

Fig. 1

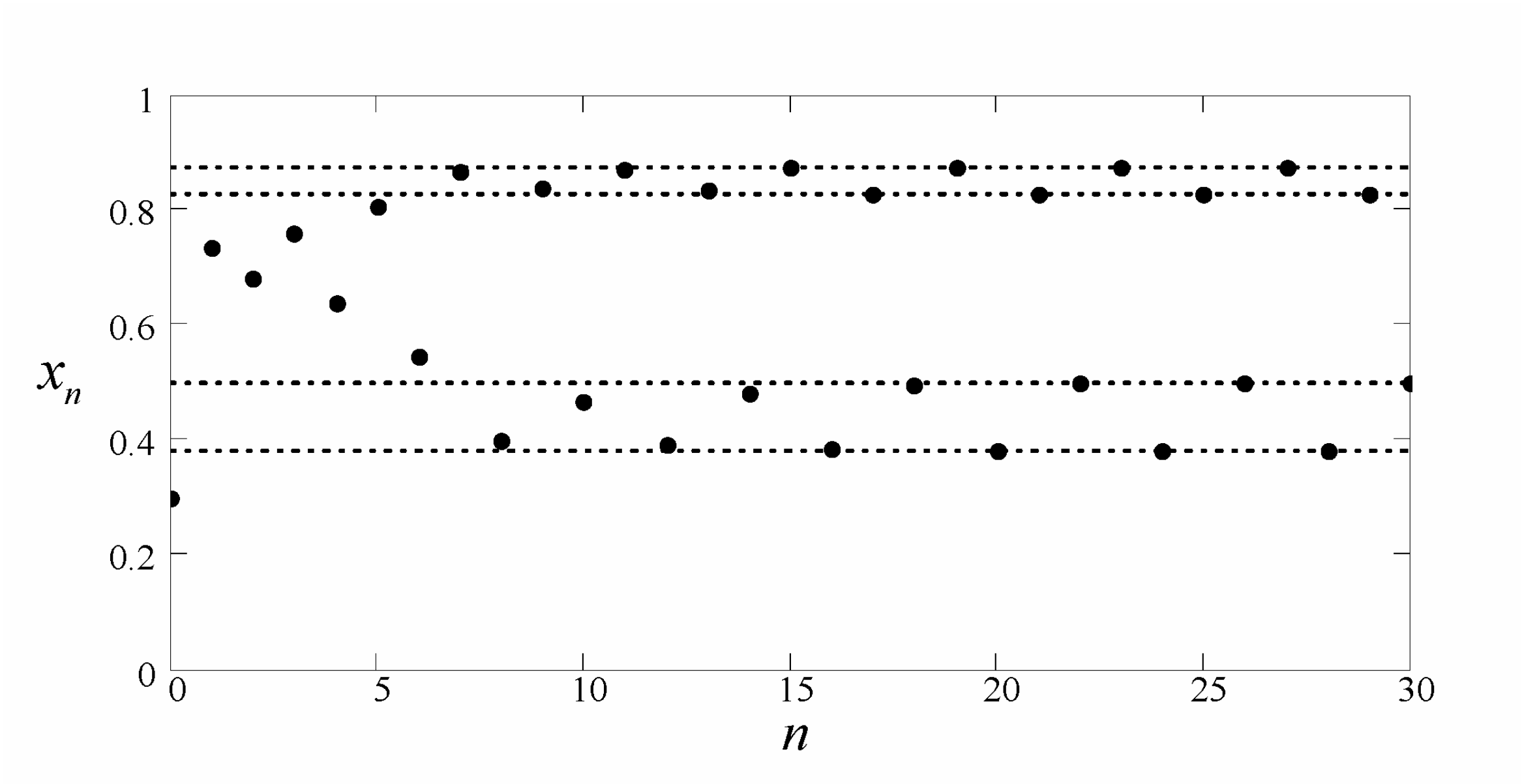

Fig. 2

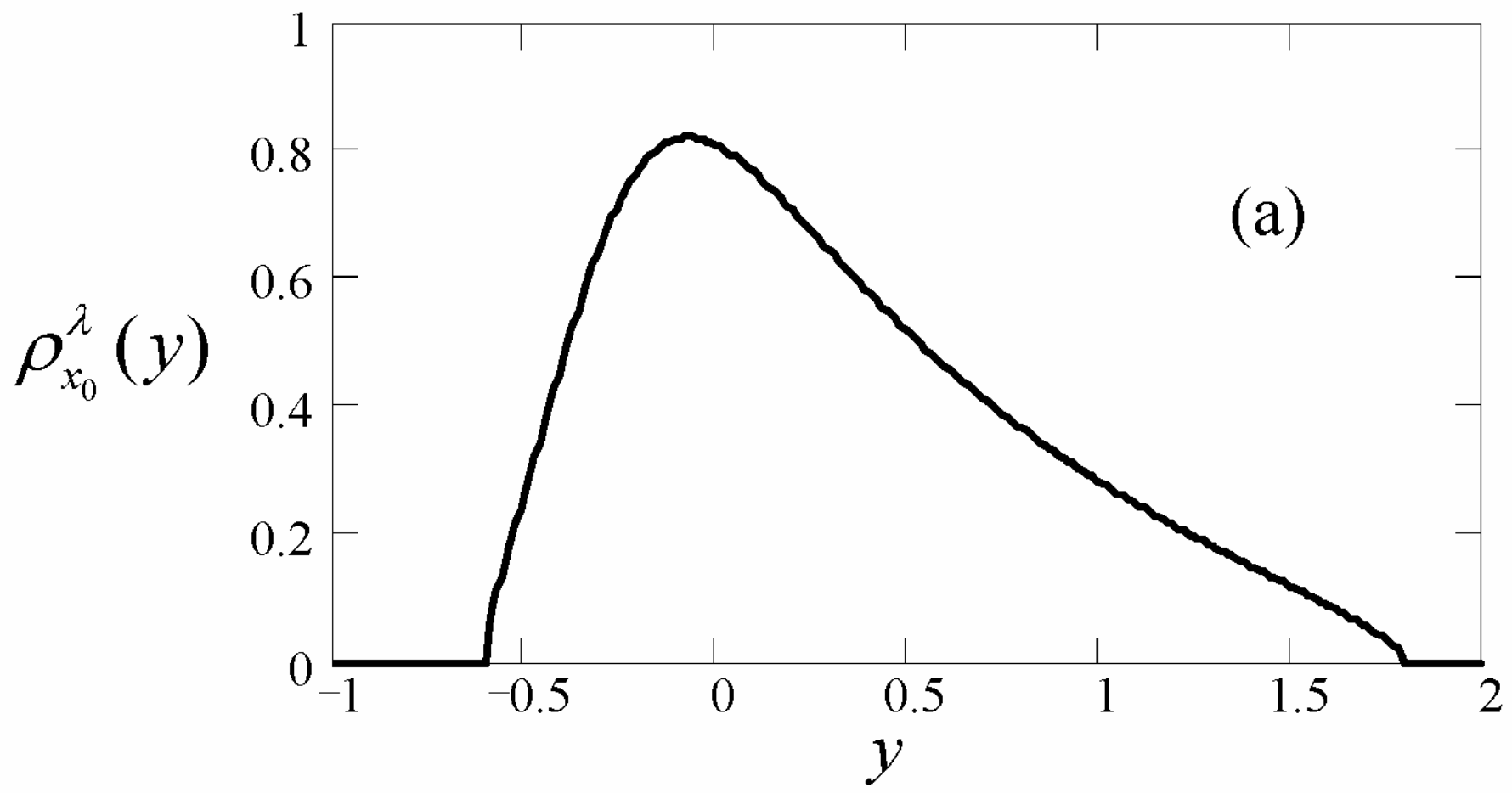

Fig. 3a

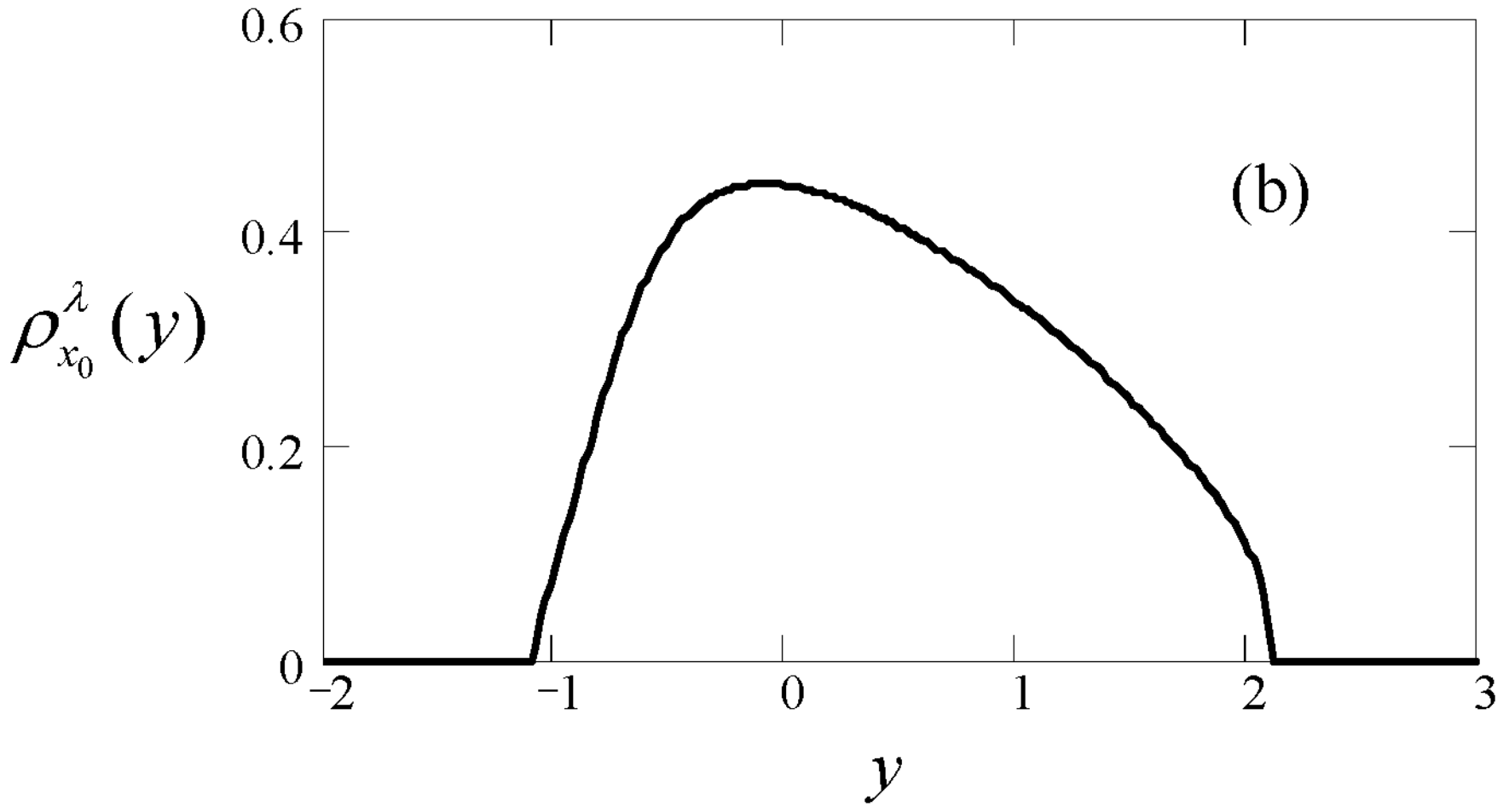

Fig. 3b

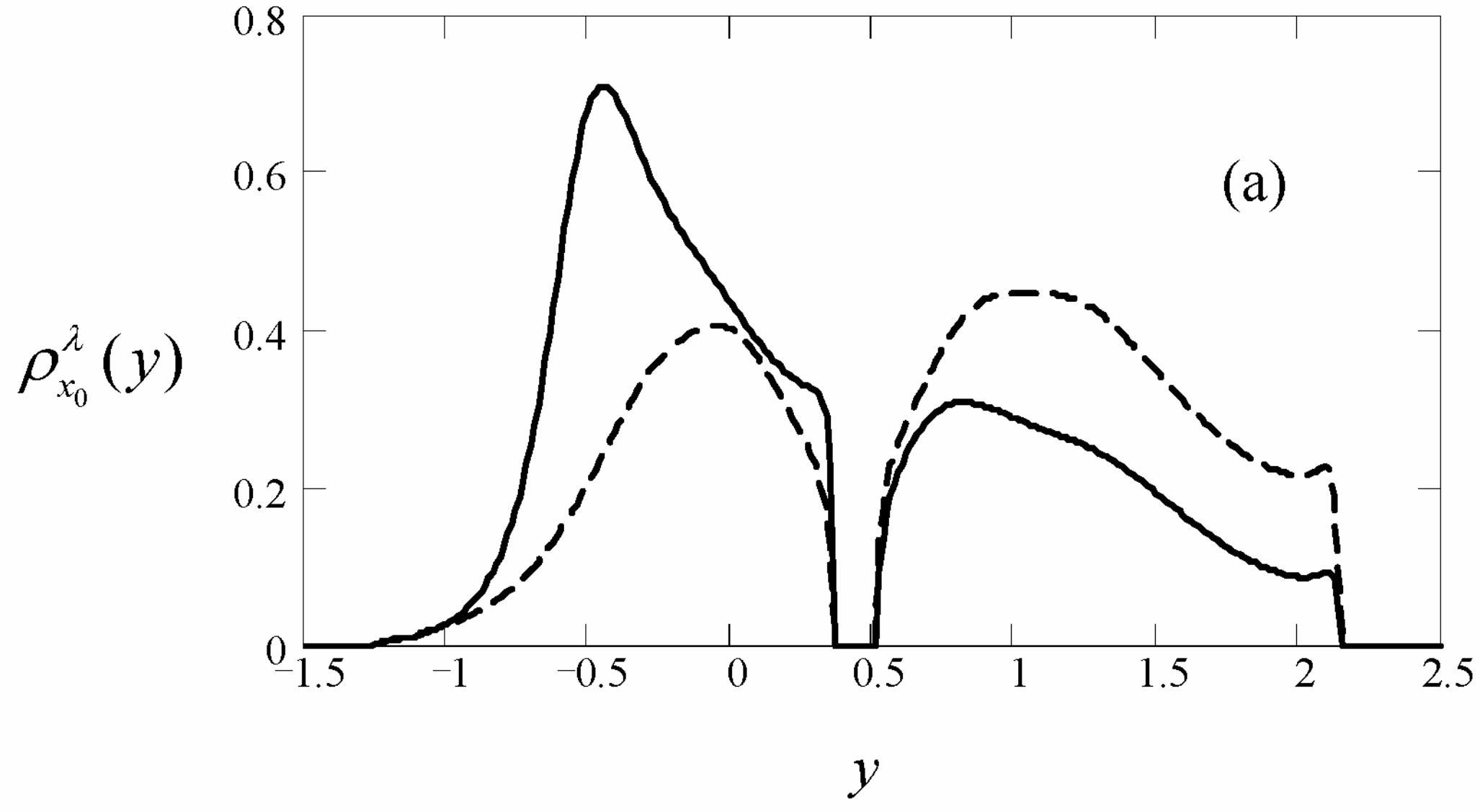

Fig. 4a

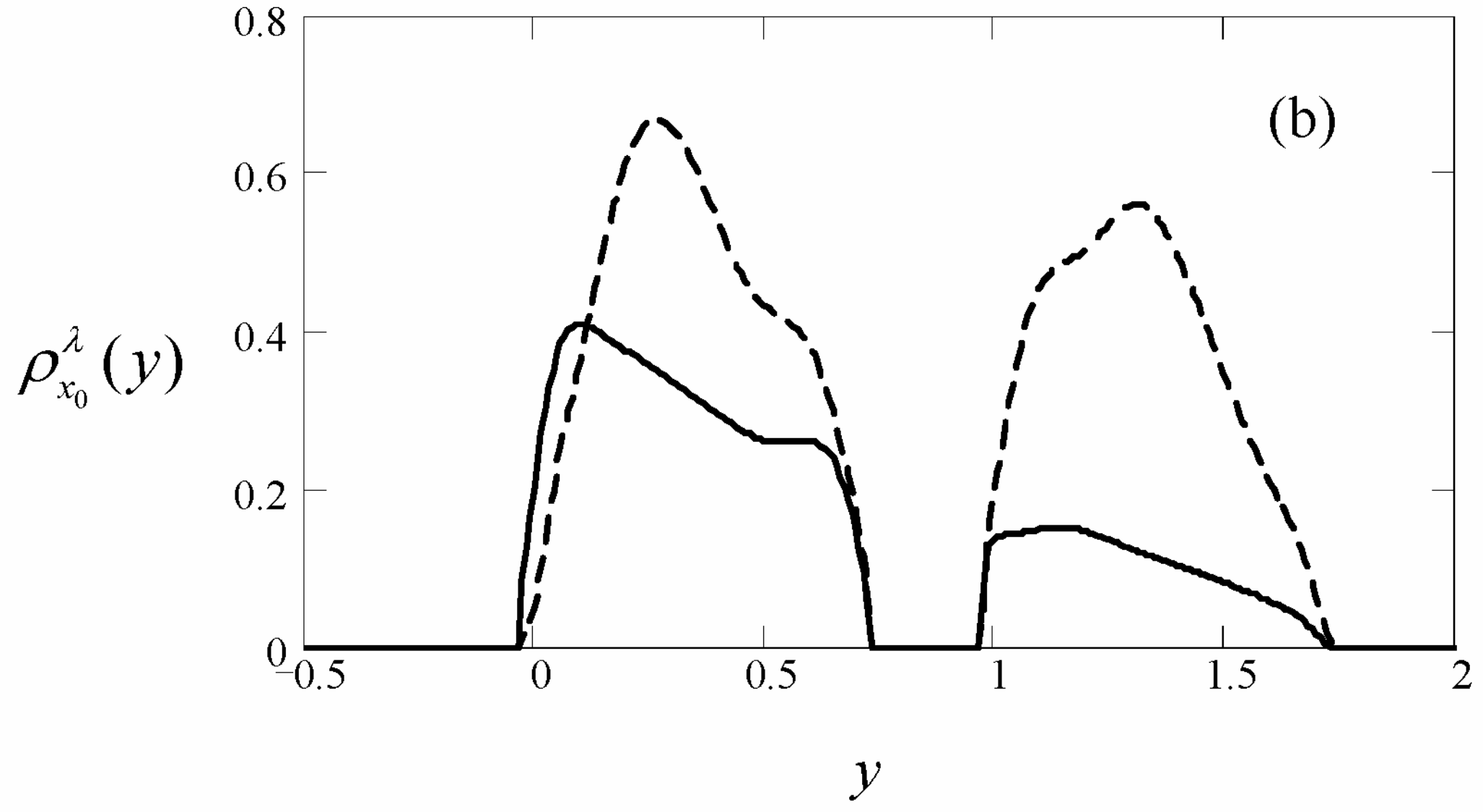

Fig. 4b

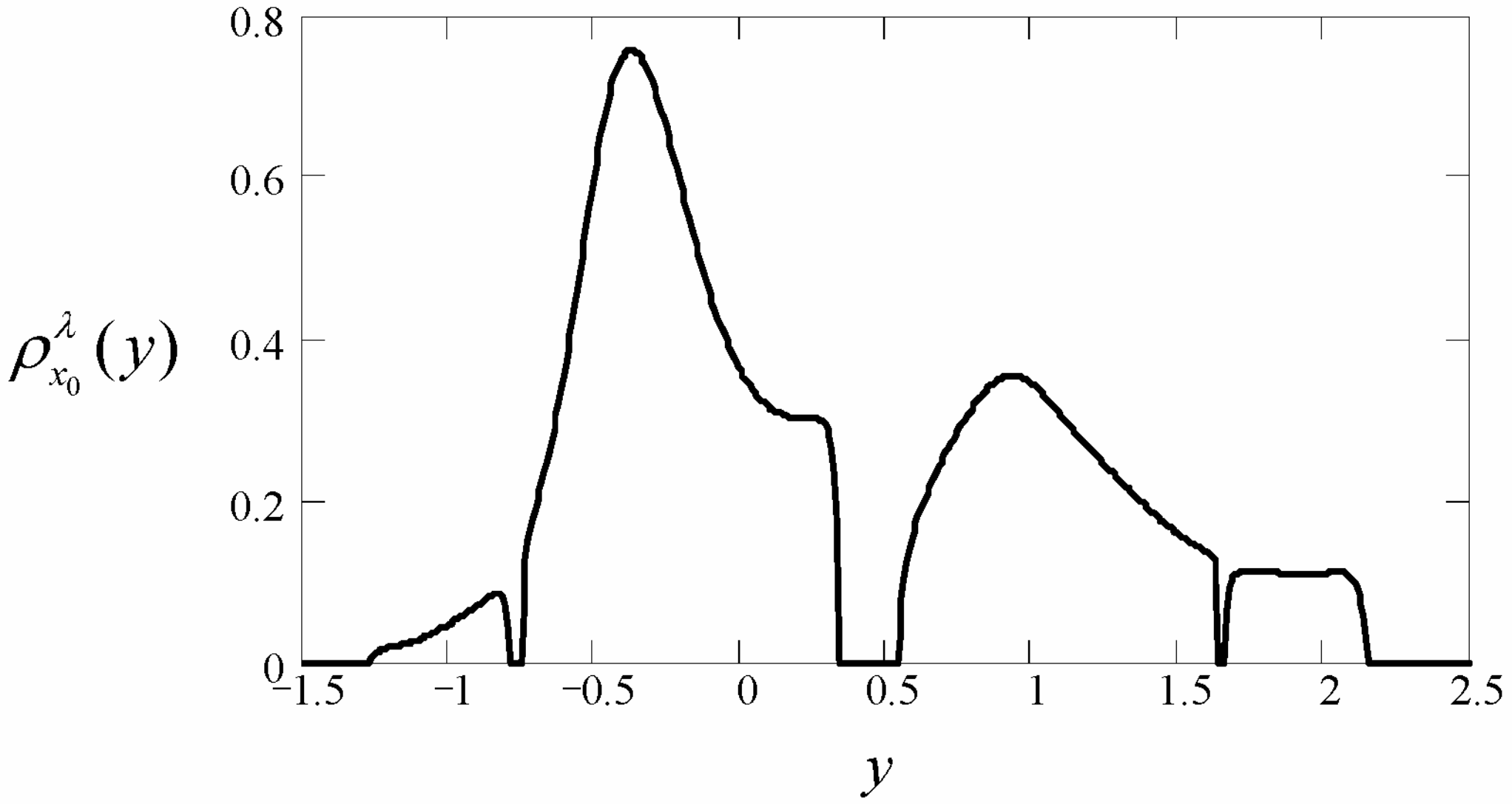
0.8
0.6
0.4
0.2
0
$\rho_{x_0}^{\lambda}(y)$
−1.5 −1 −0.5 0 0.5 1 1.5 2 2.5
$y$

Fig. 5

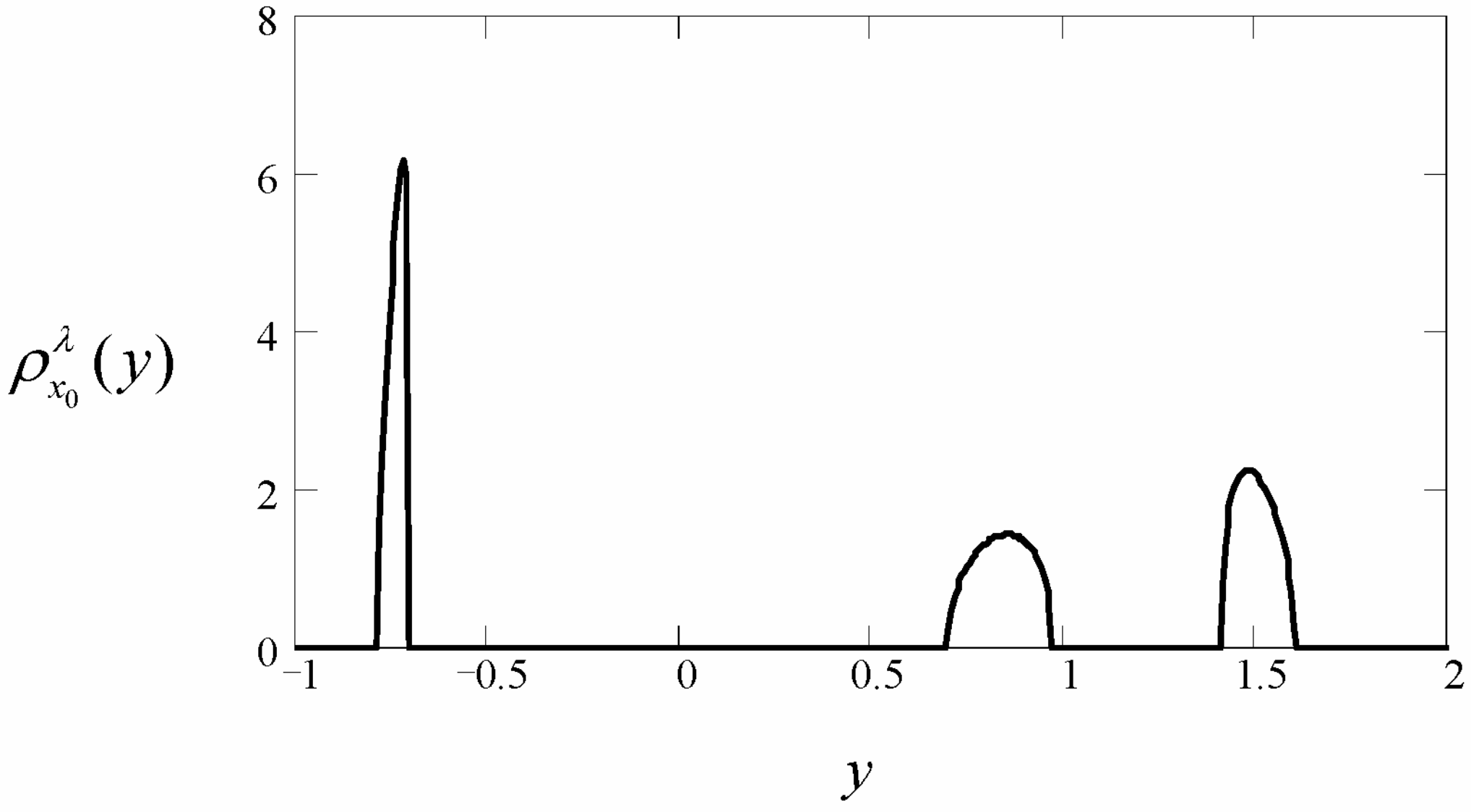

Fig. 6

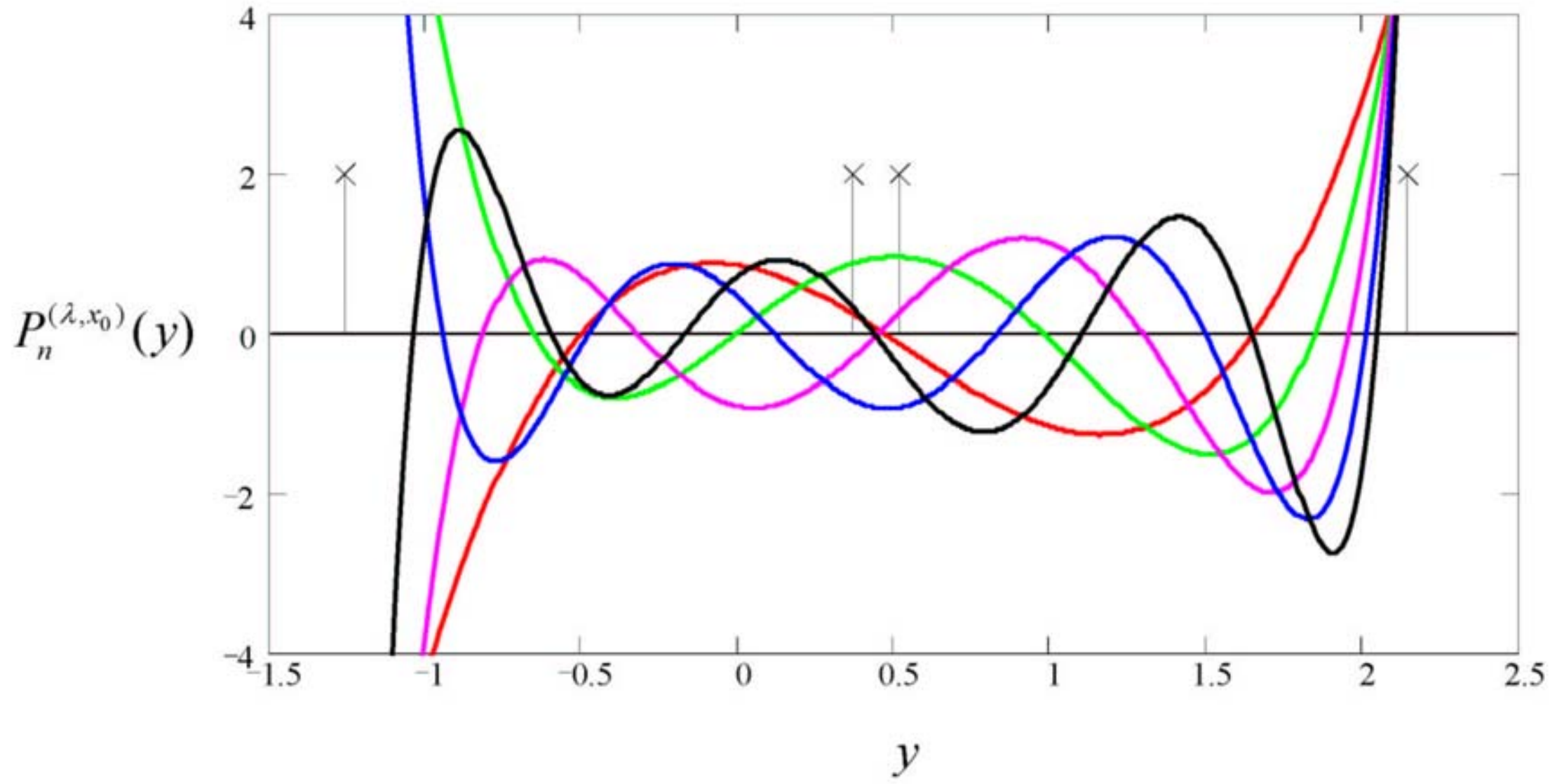

Fig. 7